\documentclass[pra,twocolumn,aps,floatfix]{revtex4}
\usepackage[pdftex]{graphicx}

\begin{document}

\title{A MULTISECTION BROADBAND IMPEDANCE TRANSFORMING BRANCH-LINE HYBRID}

\author{S. Kumar, C. Tannous}
\altaffiliation{Present address: Laboratoire de Magnétisme de Bretagne, UPRES A CNRS 6135, Université de Bretagne Occidentale, BP: 809 Brest CEDEX, 29285 FRANCE}
\author{T. Danshin}
\affiliation{Electrical Engineering, University of Saskatchewan, \\
TRLabs, Suite 108, 15 Innovation Boulevard Saskatoon SK, S7N 2X8, Canada}

\date{March 28, 2001}

\pacs{PACS numbers: 84.40.-x, 84.40.Dc, 84.30.Bv}

\begin{abstract}

Measurements  and  design  equations  for  a  two   section
impedance   transforming   hybrid   suitable    for    MMIC
applications and a new method of synthesis for multisection
branch-line  hybrids  are reported.  The  synthesis  method
allows  the  response to be specified either of Butterworth
or  Chebyshev  type. Both symmetric (with equal  input  and
output    impedances)    and    non-symmetric    (impedance
transforming) designs are feasible. Starting from  a  given
number   of  sections,  type  of  response,  and  impedance
transformation ratio and for a specified midband  coupling,
power  division  ratio,  isolation  or  directivity  ripple
bandwidth,  the set of constants needed for the  evaluation
of the reflection coefficient response is first calculated.
The  latter is used to define a driving point impedance  of
the  circuit,  synthesize it and  obtain  the  branch  line
immittances  with the use of the concept of  double  length
unit  elements  (DLUE). The experimental  results  obtained
with microstrip hybrids constructed to test the validity of
the  brute  force optimization and the synthesized  designs
show very close agreement with the computed responses.\\

{\bf Keywords}: Microwave circuits, distributed parameter synthesis,
Butterworth and Chebyshev filters.

\end{abstract}

\maketitle

\section{Introduction}
Branch line hybrids are extensively used in the realization
of  a  variety of microwave circuits. Balanced mixers, data
modulators,  phase shifters, and power combined  amplifiers
are  some examples of such circuits. Single section hybrids
have  a  limited  bandwidth. For example, a single  section
quad  hybrid  with equal power division has a bandwidth  of
about  $15 \%$ over which the power balance is within 0.5  dB.
It  is  well  known  that the operating  bandwidth  can  be
greatly   increased   using  multisection   hybrids.   Most
applications require also that the $50 \Omega$ input impedance  be
transformed  to a higher or lower impedance. A hybrid  with
built-in  impedance  transformation  is  limited   by   the
practical  realizability  of the  line  impedances  of  the
various branches.\\

Although a higher bandwidth may be achieved using a coupled
line  configuration instead of a branch line  one,  coupled
line  hybrids  are  difficult to realize,  particularly  if
microwave    monolithic    integrated    circuits    (MMIC)
implementation  is  used. The branch line  hybrid  has  the
advantage that it may be realized using slot lines  in  the
ground  plane  of a microstrip circuit. In  this  case  the
hybrid requires virtually no additional real estate on  the
chip.  This  may  be  an important consideration  when  the
hybrid is part of a larger MMIC circuit. At lower microwave
frequencies  (5  GHz or less) a lumped element  realization
similar  to  that of [1] may be used to implement  an  MMIC
hybrid.\\

For  some applications it may be sufficient to employ a two
section  impedance  transforming  hybrid  which  has  ideal
performance  at  the center of the desired frequency  band.
Design equations for such a hybrid are derived in the  next
section.   A  general  synthesis  method  for  multisection
hybrids  is also reported. The hybrid is in effect  a  four
port impedance transforming structure. Synthesis procedures
for  two  port  impedance transformers using  quarter  wave
sections  to  realize  a  Butterworth  or  Chebyshev   type
response  are  well  known. The synthesis  method  reported
here, similar to that used by Levy and Lind [6], is applied
to  the  hybrid  with only the two port even  mode  circuit
being synthesized. There are however, important differences
from [6] which are brought out in the section on synthesis.
This  paper  is  organized as follows:  In  section  II  we
describe  and  analyze the performance of  the  two-section
broadband  hybrid and in section III a general  method  for
the synthesis of multisection hybrids is described. Section
IV   contains   a   comparative  discussion   between   the
measurements, optimization and synthesis and  contains  our
conclusions.

\section{ANALYSIS AND PERFORMANCE OF THE TWO-SECTION HYBRID}

A  two-section branch-line quadrature hybrid is shown in Figure 1.
Using odd-even mode analysis, the even and odd mode cascade
element matrices at the center frequency are given by:

\begin{eqnarray}
&& M_{e}= \left[ \begin{array}{ c c}
A & B \\
C & D 
\end{array}  \right] \mbox {; } M_{o}= \left[ \begin{array}{ c c}
A & -B \\
-C & D 
\end{array}  \right]   \nonumber \\
&& \mbox{where:}    \nonumber \\
&& A= \frac{b^2}{cd} -1; B=-j \frac{b^2}{c}; C=-j( \frac{1}{a}
+\frac{1}{d}- \frac{b^2}{acd})   \nonumber \\ 
&&   \hspace{2cm} \mbox{ and } D= \frac{b^2}{ac}-1
\end{eqnarray}

Use  of  the  relation between cascade parameters  and  the
reflection   and   transmission   coefficients   [3]    and
application  of  matching and perfect isolation  conditions
($S_{11}= 0$ and $S_{41}= 0$ at the center frequency) results in:

\begin{equation}
A=DZ_{01} /Z_{02},   B=  C Z_{01}Z_{02}
\end{equation}

where  $Z_{01}, Z_{02}$ are input (ports 1 and 4) and output (ports
2 and 3) impedances respectively.\\

\begin{figure}[htbp]
\centering{\includegraphics[width=2.5in,angle=0]{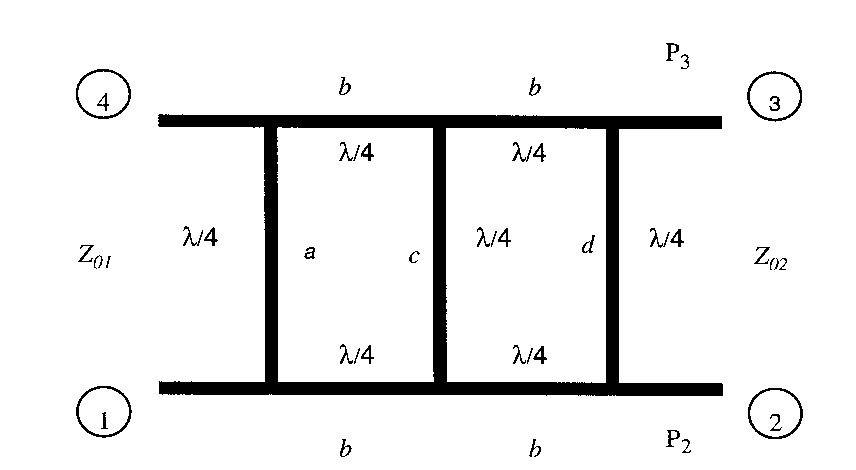}}
\caption{Two-section branch line impedance  transforming
hybrid $a, b, c, d$ are the characteristic impedances of  the
quarter wave branches.}
\label{fig1}
\end{figure}

The  normalized voltage waves $b_{2}$ and $b_{3}$ at the output ports
are  given  in  terms of the odd and even mode transmission
coefficient. $T_{e} \mbox{ and }T_{o}$ by $b_{2}= (T_{e}+T_{o})/2$ and $b_{3}= (T_{e}-T_{o})/2$. Once  again  using  the relation between  the  transmission
coefficients and the cascade parameters [3] the output port
power ratio may be written as:

\begin{equation}
\frac{ P_{3} }{ P_{2} }  =  \frac{ P_{out} \mbox{ at } port3 }{ P_{out} \mbox{ at } port2 }=k^2
\end{equation}
with:
\begin{equation}
 k  = | \frac{b_{3} }{ b_{2} }| = | \frac{C}{A} Z_{01} |
\end{equation}
                                       
This gives:
\begin{equation}
C^{2}=-k^{2}A^{2}/{Z_{01}}^{2}                 
\end{equation}

The  negative  sign was retained in (5) as  equation  (1)
implies  $C^2$  to be a negative quantity. A relation  between
the  line impedances a  and d is found by first solving for
$b^{2}/c $  using  (1),  the  two components  of  (2)  and  then
equating the two values of $b^{2}/c $  thus found. This gives:

\begin{equation}
\frac{1}{a^{2}}=\frac{r}{d^{2}}+\frac{(1-r)}{r{Z_{01}}^{2}}
\end{equation}

where  $r$, the impedance transformation ratio is defined  by
the ratio $Z_{01} /Z_{02}$.
In order to obtain design equations for the line impedances
$a,  d$ another equation relating these impedances is needed.
This  is  obtained by applying the losslessness  condition:
$|b_{2}|^{2}+|b_{3}|^{2}=1$ along with (3):

\begin{equation}
|b_{2}|^{2}=1/(1+k^{2})
\end{equation}                                  

Substitution for $b_{2}$ in terms of $T_{e} \mbox{ and }T_{o}$ and use of (1)
and (5) result in:

\begin{equation}
 b_{2} = \frac{1}{A \sqrt{r} (1+k^2)}
\end{equation}                                        

From (7) and (8):

\begin{equation}
 A = - \frac{1}{ \sqrt{r (1+k^2)}}
\end{equation}

The negative sign was retained in equation (9) as this
solution gives non negative values of the branch
impedances.
Using (9) and relations between the cascade parameters and
line impedances in (1) we can obtain the second equation
relating $a$ and $d $ as:

\begin{equation}
 a = d \frac{ \sqrt{r (1+k^{2})} -1  }{ \sqrt{r (1+k^{2})} -r}
\end{equation}

From [6] and [10] the line impedance  $a$ is:

\begin{eqnarray}
 a & = &  Z_{01} \frac{ \sqrt{r (t^{2}-r) } }{ t -r} \hspace{1cm} \mbox{ with $t$ defined as } \nonumber \\
 t & =  & r \sqrt{1+k^2}
\end{eqnarray}

Substitution back gives:

\begin{equation}
 d = Z_{01} \frac{ \sqrt{r (t^{2}-r) } }{ t -1} 
\end{equation}
and:

\begin{equation}
 b^{2}/c= Z_{01} \sqrt{r -  r^{2}/t^{2}} 
\end{equation}

Equations (11), (12) and (13) can be used to design the two
section hybrid with a given impedance transformation  ratio
$r$  and  the power ratio (coupling) $k^2$. Note that the  ratio
$b/c$  can  be  chosen to be different from 1.  However,  $b=c$
gives  maximum bandwidth when the best performance  at  the
band center is specified. Impedances $ b$  and $c $ are commonly
chosen to be equal.\\

\begin{figure}[htbp]
\centering{\includegraphics[width=2.5in,angle=0]{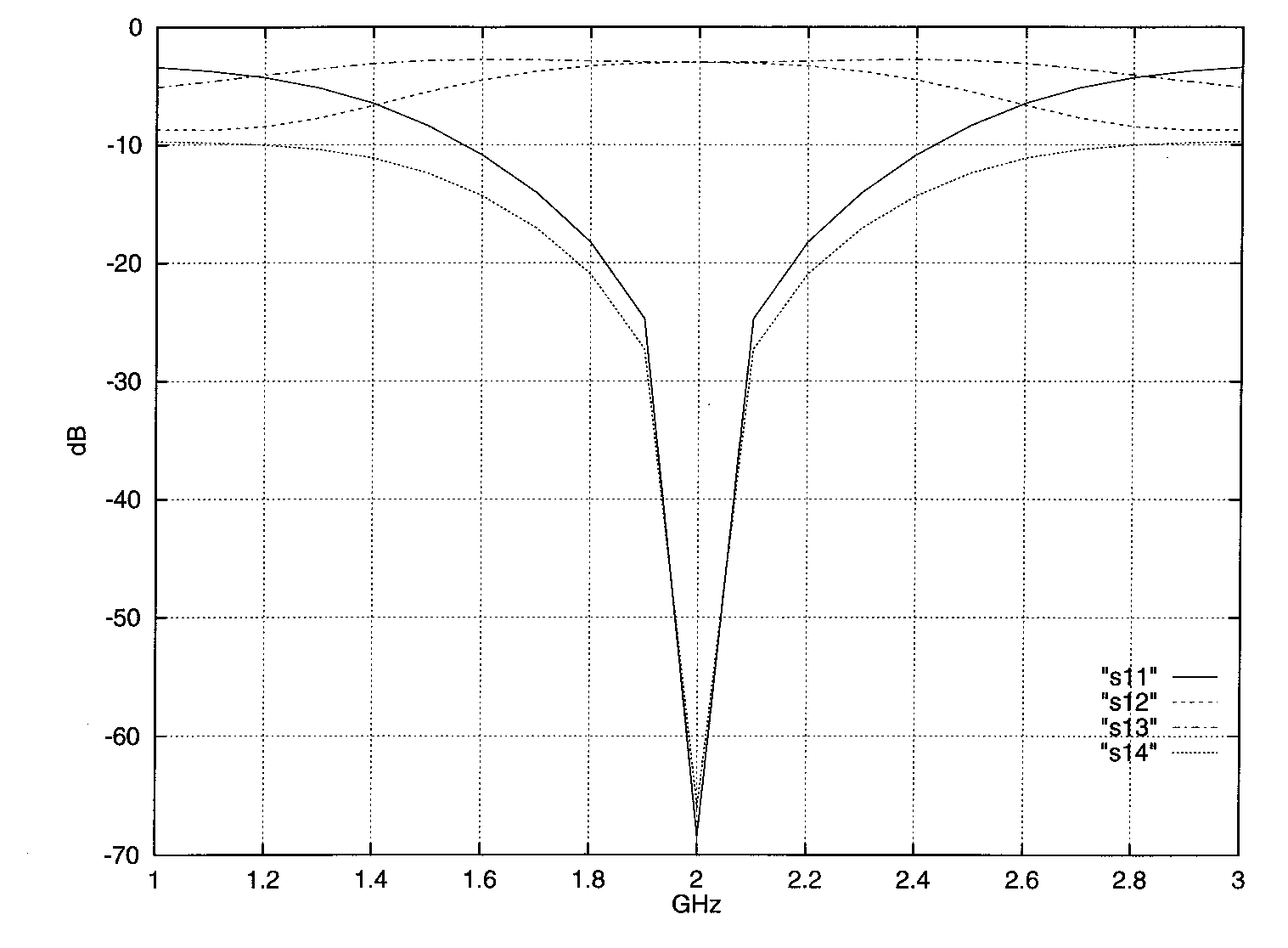}}
\caption{Computed frequency response of a two-section  $50\Omega$
to  $35\Omega$  hybrid.  The impedance values  used  are  $a=72.5\Omega,
b=29.6\Omega, c=29.6\Omega$ and $d=191.25\Omega$.}
\label{fig2}
\end{figure}

For  equal  power  division, $k=1$ and $t^{2}= 2r^{2}$.  The  minimum
value for $r$  for non negative branch impedances is 0.5.  In
practice,  $r$  in  the range of .7 to 1.3 for  a  50$\Omega$ input
impedance  gives  practically realizable  line  impedances.
Referring  to Fig.1, the computed line impedance values  of
an  equal power division, 50 to 35$\Omega$ two-section hybrid are:
$a=72.5\Omega$, $b=29.6\Omega$, $c=29.6\Omega$ and $d=191.25\Omega$.

\begin{figure}[htbp]
\centering{\includegraphics[width=2.5in,angle=0]{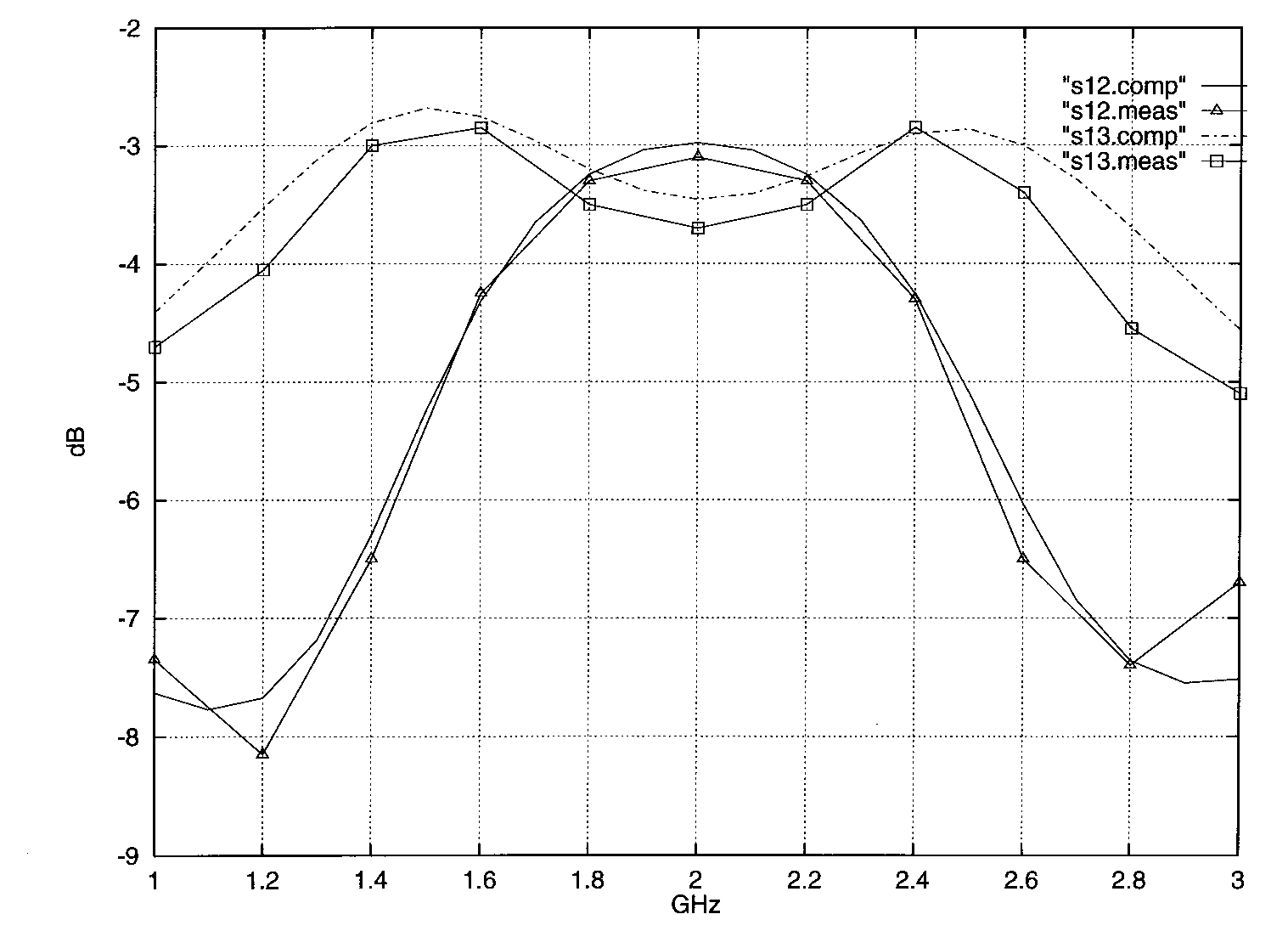}}
\caption{ Computed and measured $S_{12}$ and $S_{13}$ responses  for
the optimized hybrid. The hybrid was fabricated on a Rogers
5880,  0.031 inch thick Duroid substrate. A wide band three-
section Chebyshev transformer was used at the output  ports
to   transform  the  $35\Omega$  impedance  back  to  $50\Omega$  for   the
measurement.  The optimized impedance values  are  $a=90  \Omega$,
$b=39 \Omega$, $c=56 \Omega$ and $d=110\Omega$.}
\label{fig3}
\end{figure}

\begin{figure}[htbp]
\centering{\includegraphics[width=2.5in,angle=0]{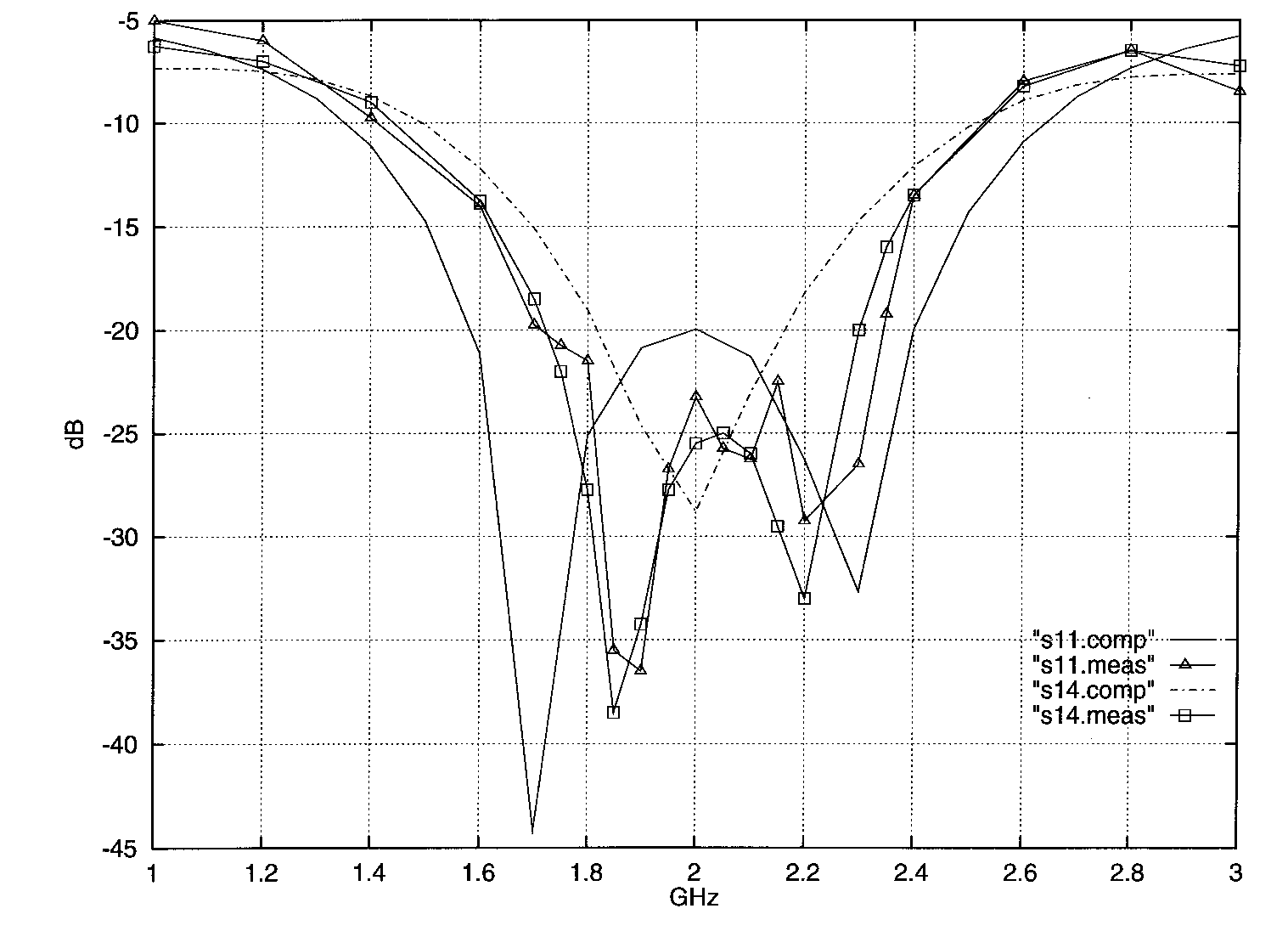}}
\caption{Computed and measured $S_{11}$ and $S_{14}$ responses  for
the optimized hybrid described in Fig.3.}
\label{fig4}
\end{figure}

\begin{figure}[htbp]
\centering{\includegraphics[width=2.5in,angle=0]{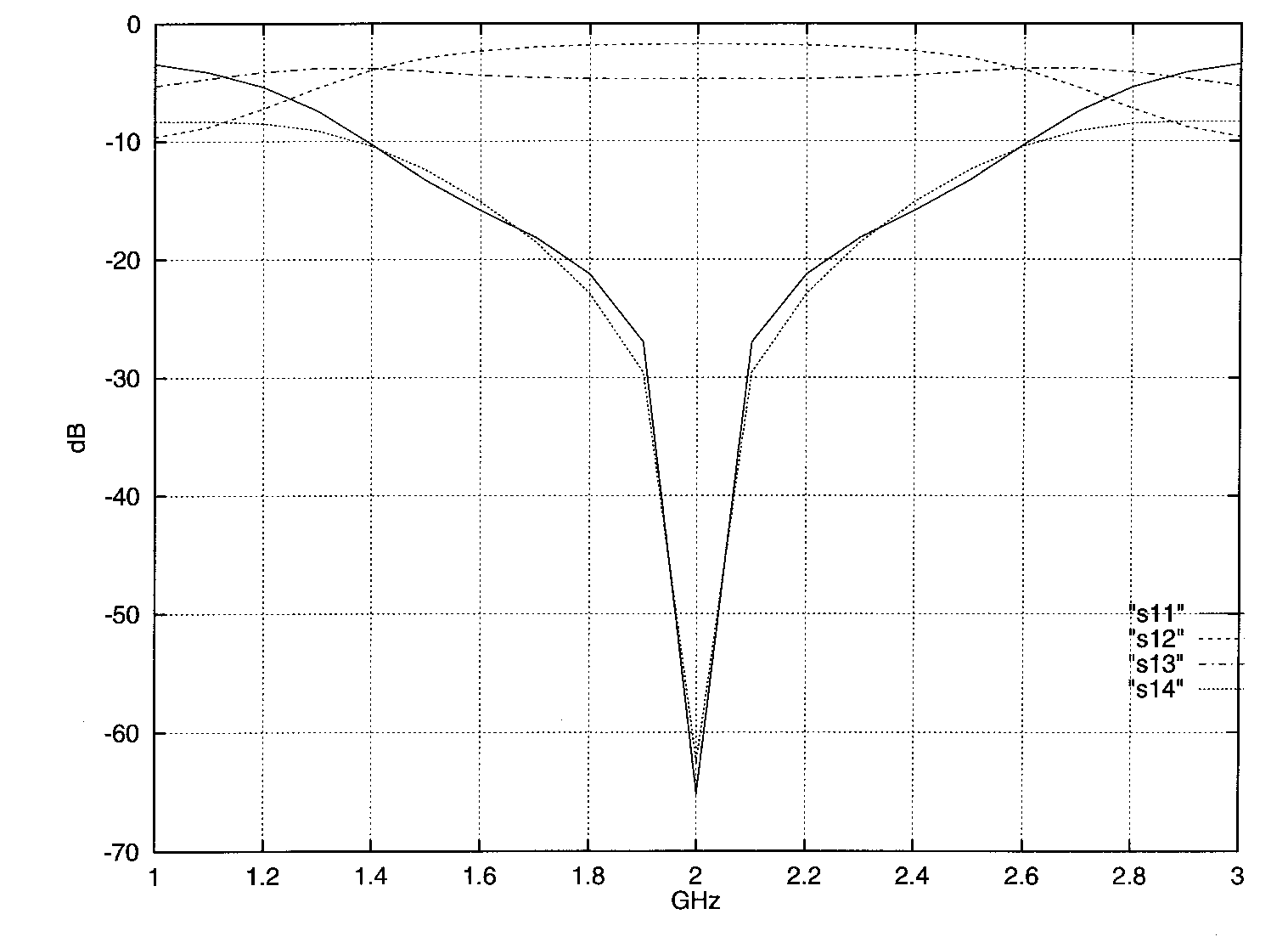}}
\caption{Computed  response  of  a  two-section  $50\Omega$  to
$50\Omega$  with  3 dB unequal power division hybrid. The computed
values  for a 3 dB unequal power division, i.e.  $k=  0.707$,
non impedance transforming hybrid are $a=d=157 \Omega, b=c=29 \Omega$.}
\label{fig5}
\end{figure}

The computed frequency response for a 2 GHz hybrid is shown
in  Fig. 2. As may be seen from this figure, a 0.5dB output
balance bandwidth of $25\%$ is feasible. However, the response
can  be  further  improved  by  computer  optimization.  In
carrying  out the optimization, limits were placed  on  the
impedance  values  in order to yield an  easily  realizable
design.  A  multisection hybrid offers the  flexibility  of
carrying  out this optimization quite effectively.  The  T-
junction discontinuity effects can also be included in  the
program.  Such  effects become quite  important  at  higher
frequencies.  Referring to Fig. 1, the optimized  impedance
values  are:  $a= 90\Omega, b=39\Omega, c=56\Omega, d=110\Omega$. 
The  hybrid  was fabricated  on  a  Rogers  5880, 0.031 inch  thick  Duroid
substrate.  A wide band three section Chebyshev transformer
was used at the output ports to transform the $35\Omega $ impedance
back  to $50\Omega$ for the measurement. The computed and measured
results  for $S_{12}$ and $S_{13}$ are shown in Fig.3 while the  same
for $ S_{11}$ and $S_{14}$ are shown in Fig.4. These results show that
the  agreement between the measured and computed  responses
is  quite  close and a 0.5dB balance bandwidth of  $30\% $  was
realized with a built-in impedance transformation  from  50
to 35$\Omega$.\\

\begin{figure}[htbp]
\centering{\includegraphics[width=2.5in,angle=0]{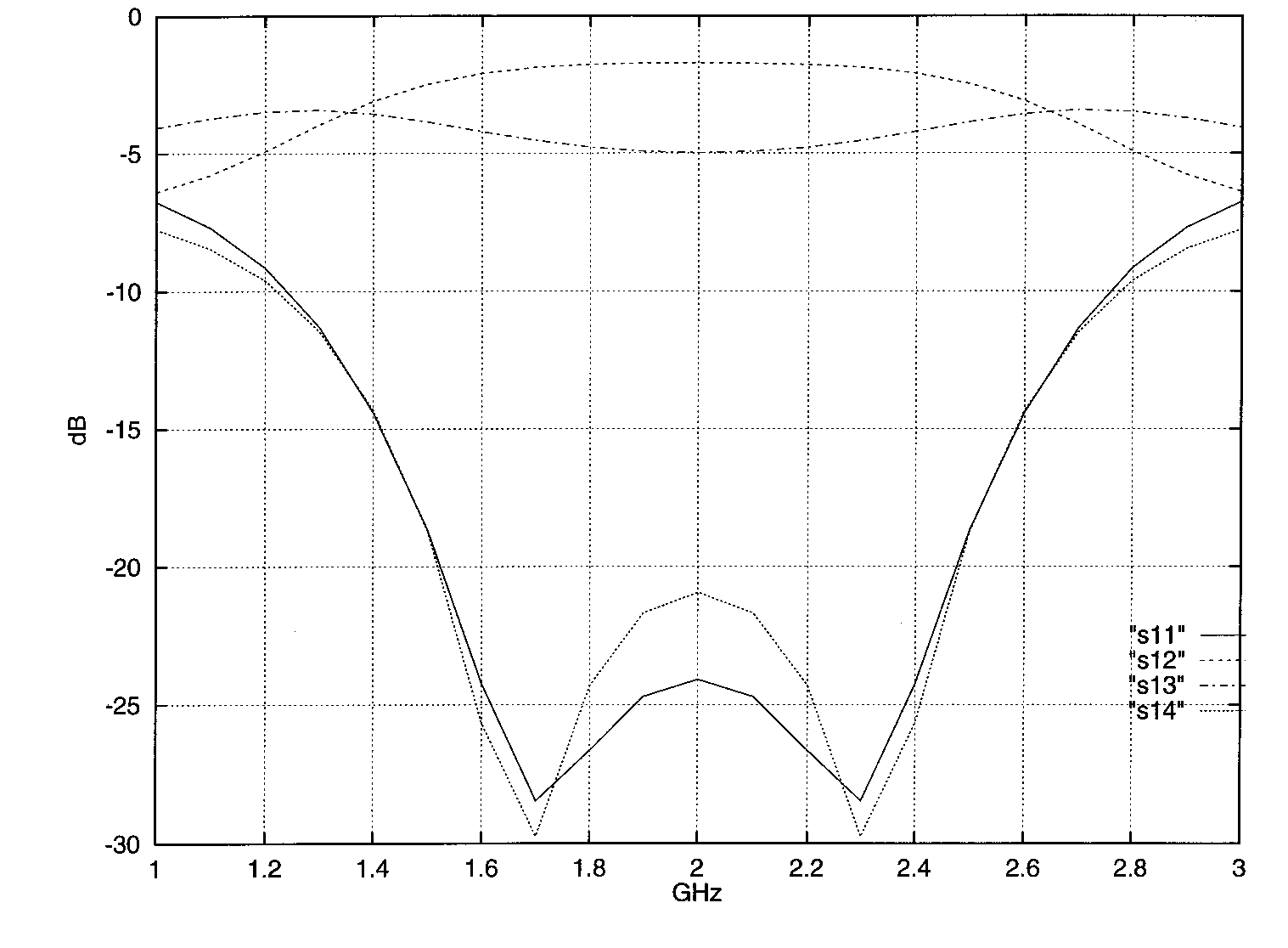}}
\caption{Computed response for the optimized 3dB  unequal
power  division hybrid ($r=1$) The optimized impedance values
are $ a=135  \Omega, b=46 \Omega, c=92 \Omega$ and $d=134 \Omega$. These  impedance
values  are  suitable  for  slotline  implementation.   The
computed  frequency  response for the optimized  impedances
has a wider bandwidth.}
\label{fig6}
\end{figure}

\begin{figure}[htbp]
\centering{\includegraphics[width=2.5in,angle=0]{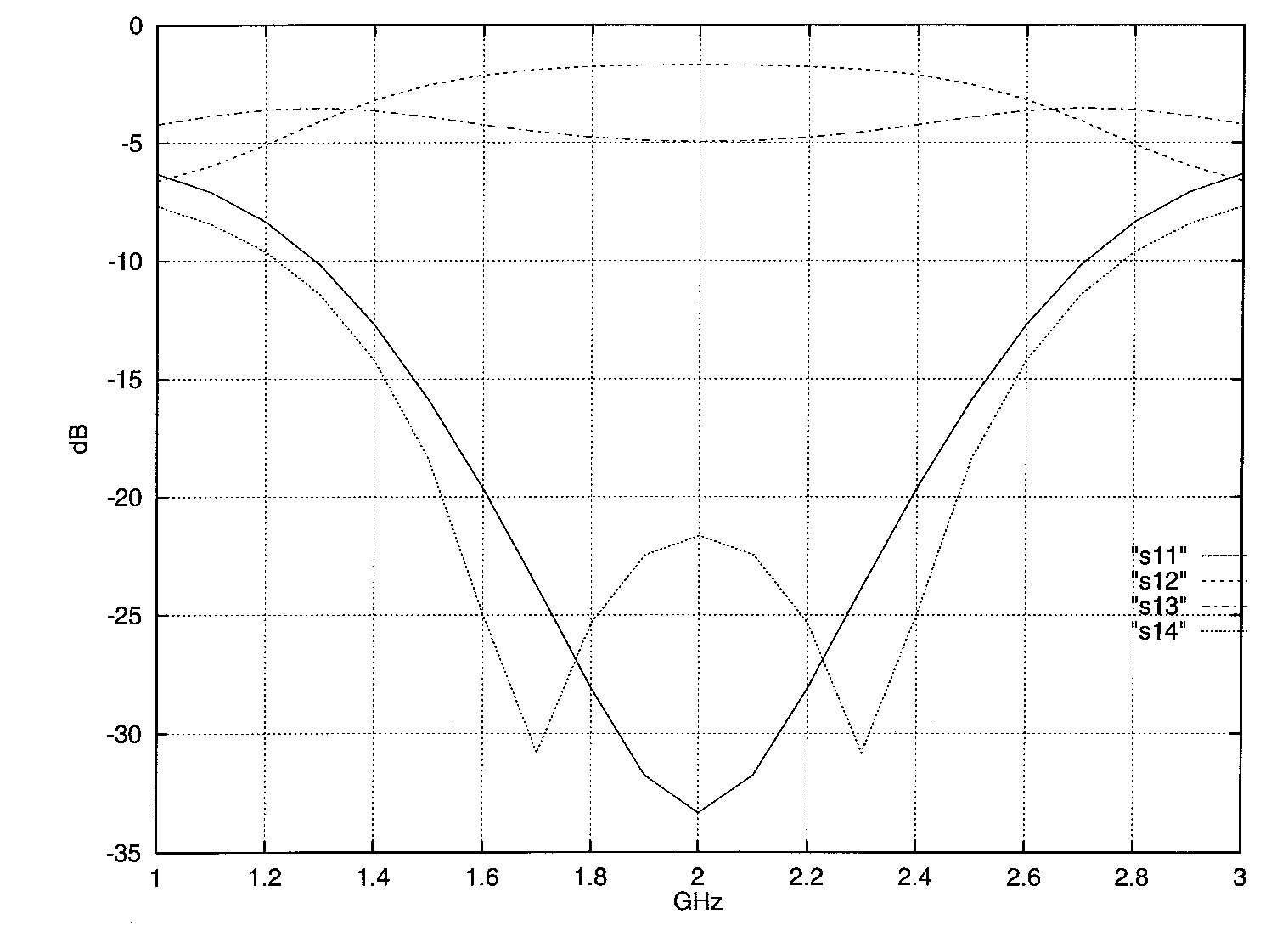}}
\caption{Computed response for the optimized 3 dB unequal
power  division 2 GHz hybrid ($r=1.2$). The optimized  branch
line impedances are $a= 170 \Omega, b= 47 \Omega, c= 77 \Omega, d=151 \Omega$. As
may  be  seen, the 0.5 dB balance bandwidth is $30\%$ and  the
return  loss and isolation are better than 20 dB  over  the
bandwidth.}
\label{fig7}
\end{figure}

The  computed values for a 3dB unequal power division, i.e.
$k= 0.707$, non impedance transforming hybrid are $a= d= 157\Omega,
b= c=  29\Omega$.  The computed frequency response  for  such  a
hybrid  is  shown  in  Fig. 5. While the  return  loss  and
isolation values in Fig. 5 are better than 20 dB over a $25\%$
bandwidth, the branch line impedances are not suitable  for
slotline   or   microstrip  implementation.  The   circuit,
however,  can  also be improved with computer optimization.
The  optimized impedance values are:  $a=135\Omega, b=46\Omega, c=92\Omega,
d=134\Omega$.  These  impedance  values  are  suitable  for
slotline  implementation. The computed  frequency  response
for the optimized hybrid is shown in Fig. 6. As can be seen
from this figure, the hybrid performance did not degrade as
a  result of optimization for realizable branch impedances.
The  branch line impedances for a 6 dB unequal power  split
$50\Omega $ to $50\Omega$  two-section hybrid did not result in practically
achievable branch line impedances for either microstrip  or
slotline.\\

The  optimized branch line impedances for a 2  GHz  50$\Omega$  to
$60\Omega$, 3dB unequal power division hybrid are $a= 170\Omega, b= 47\Omega,
c=  77\Omega, d= 151\Omega$. Fig. 7 shows the computed response of the
hybrid  and  a 0.5dB balance bandwidth of $30\%$  with  return
loss and isolation better than 20 dB over this bandwidth.\\

As   a   result   of  computer  optimization,   substantial
improvement  was possible for both equal and unequal  power
division  cases.  This  shows  that  a  design  for   ideal
performance at the band center is not adequate when maximum
possible   bandwidth  is  required.  Moreover  the   design
equations  for  an  impedance transforming,  unequal  power
division  hybrid  become quite complex  as  the  number  of
sections  increases  beyond two. In the  next  section,  we
develop  a  general  method that  can  handle  multisection
impedance  transforming hybrids and perform  the  synthesis
numerically. The starting point in this method is based  on
the analytical approach of [6].

\section{GENERAL SYNTHESIS OF A MULTISECTION BRANCH LINE HYBRID}

A  general  multisection branch-line  hybrid  is  shown  in
Figure  8. The synthesis problem of this four-port  circuit
is  equivalent  to that of synthesizing the  two-port  even
mode circuit. Starting from a given function $\Gamma_{e}/T_{e}$
 where $\Gamma_{e}$ is  the even mode reflection coefficient of the circuit and
$T_{e}$  the even mode transmission coefficient, one can extract
a cascade of double-lengths unit elements (DLUE) and single length open circuited shunt stubs [Fig.8]. The procedure is
applicable  to  the Butterworth as well  as  the  Chebyshev
response for $\Gamma_{e}/T_{e}$.

\begin{figure}[htbp]
\centering{\includegraphics[width=2.5in,angle=0]{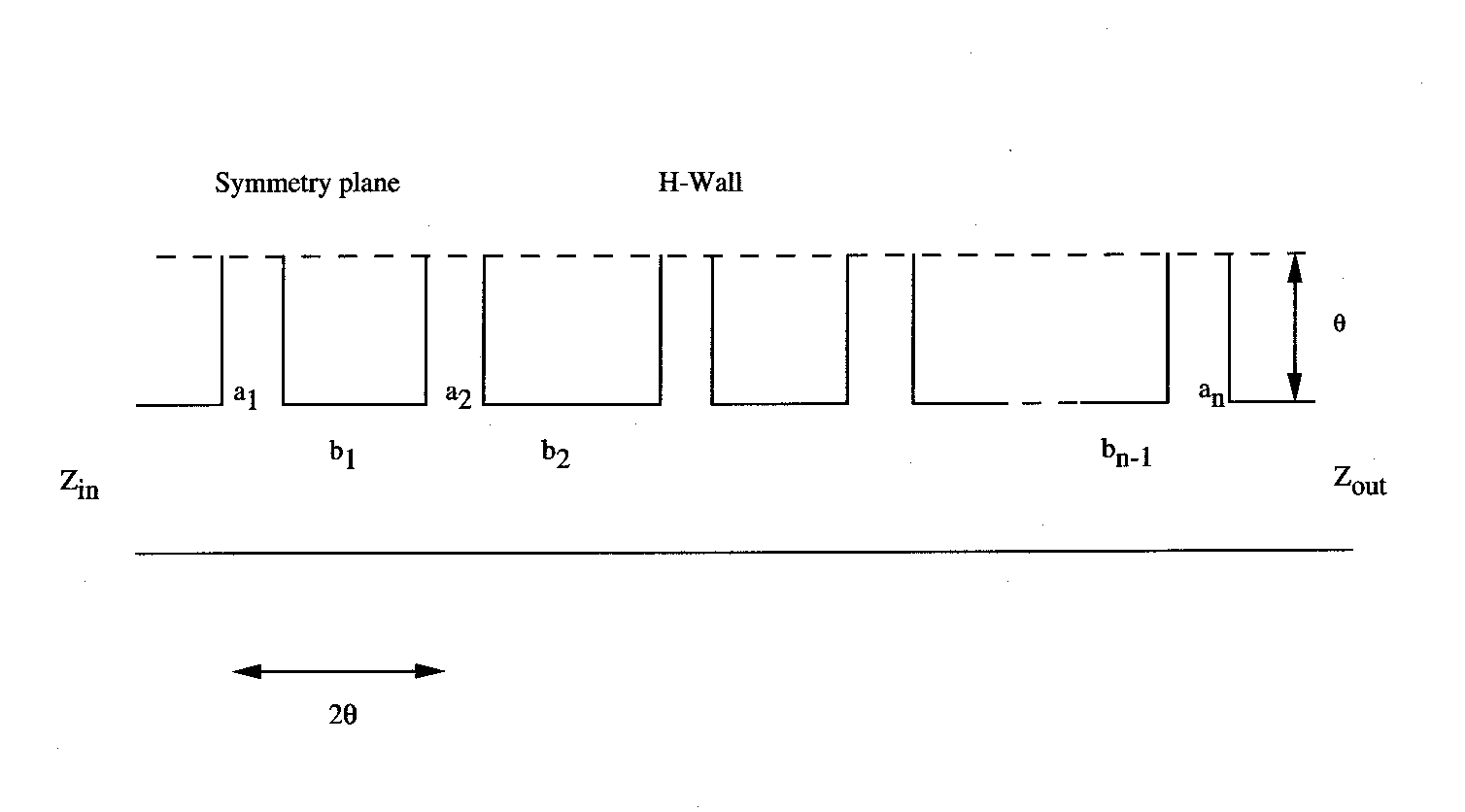}}
\caption{Even-mode circuit showing the electrical lengths
and  respective immittances. In comparison with  Fig.1  the
first immittance values are: $a_{1}= 1/a, a_{2}= 1/c, a_{3}= 1/d, b_{1}=
1/b, b_{2}= 1/b$.}
\label{fig8}
\end{figure}

\begin{figure}[htbp]
\centering{\includegraphics[width=2.5in,angle=0]{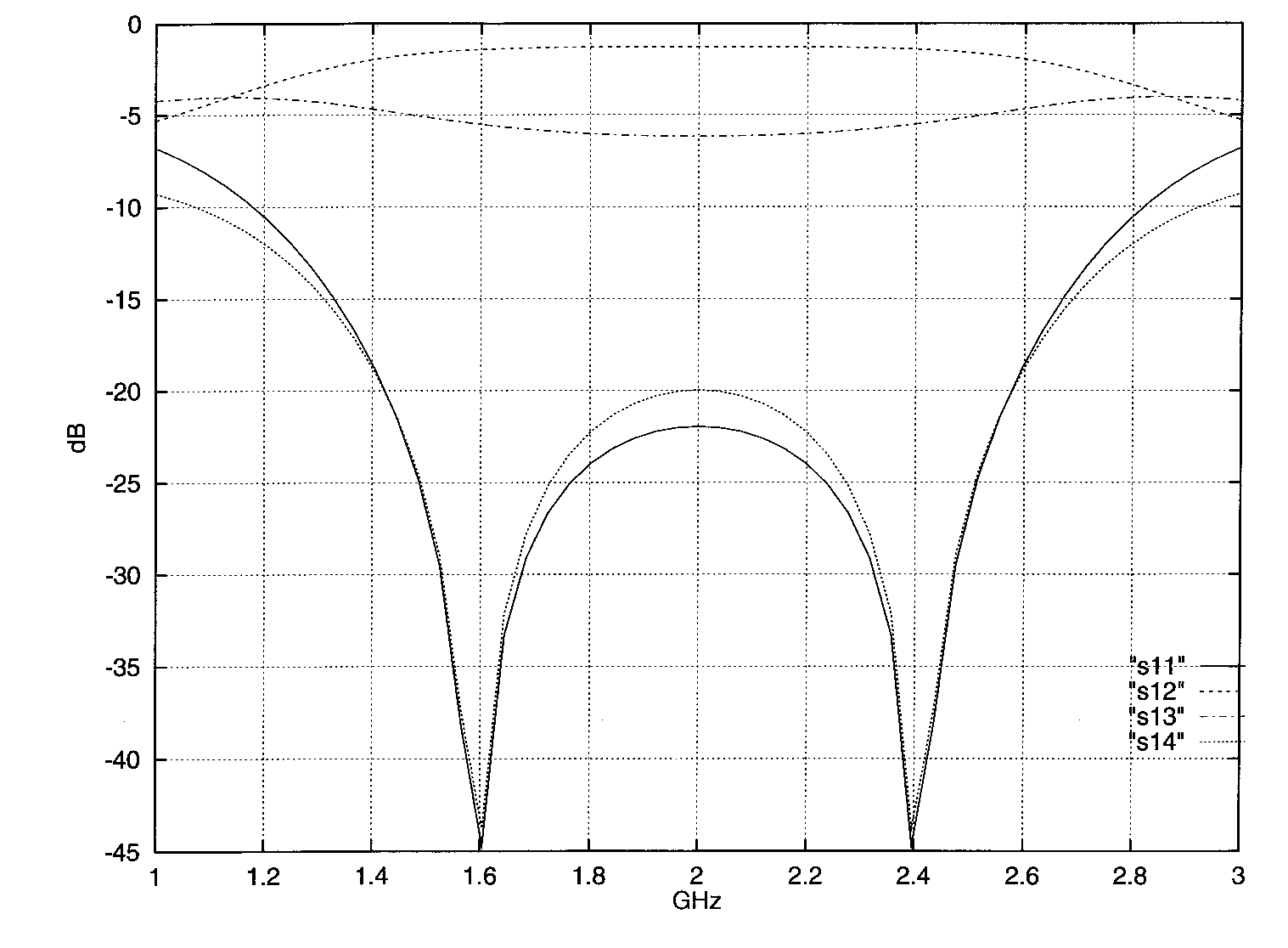}}
\caption{Computed response for the Chebyshev synthesized
hybrid of Fig. 3. The hybrid is impedance transforming  ($50
\Omega$ to $35\Omega$) and is required to have 0 dB power division ratio
at  midband  (2  GHz)  and  an isolation  of  -20  dB.  The
impedance values are: $a_{1}= 157.50 \Omega, a_{2}= 85.05 \Omega, a_{3}=  99.88
\Omega, b_{1}= 39.49 \Omega, b_{2}= 33.49 \Omega$.}
\label{fig9}
\end{figure}

\begin{figure}[htbp]
\centering{\includegraphics[width=2.5in,angle=0]{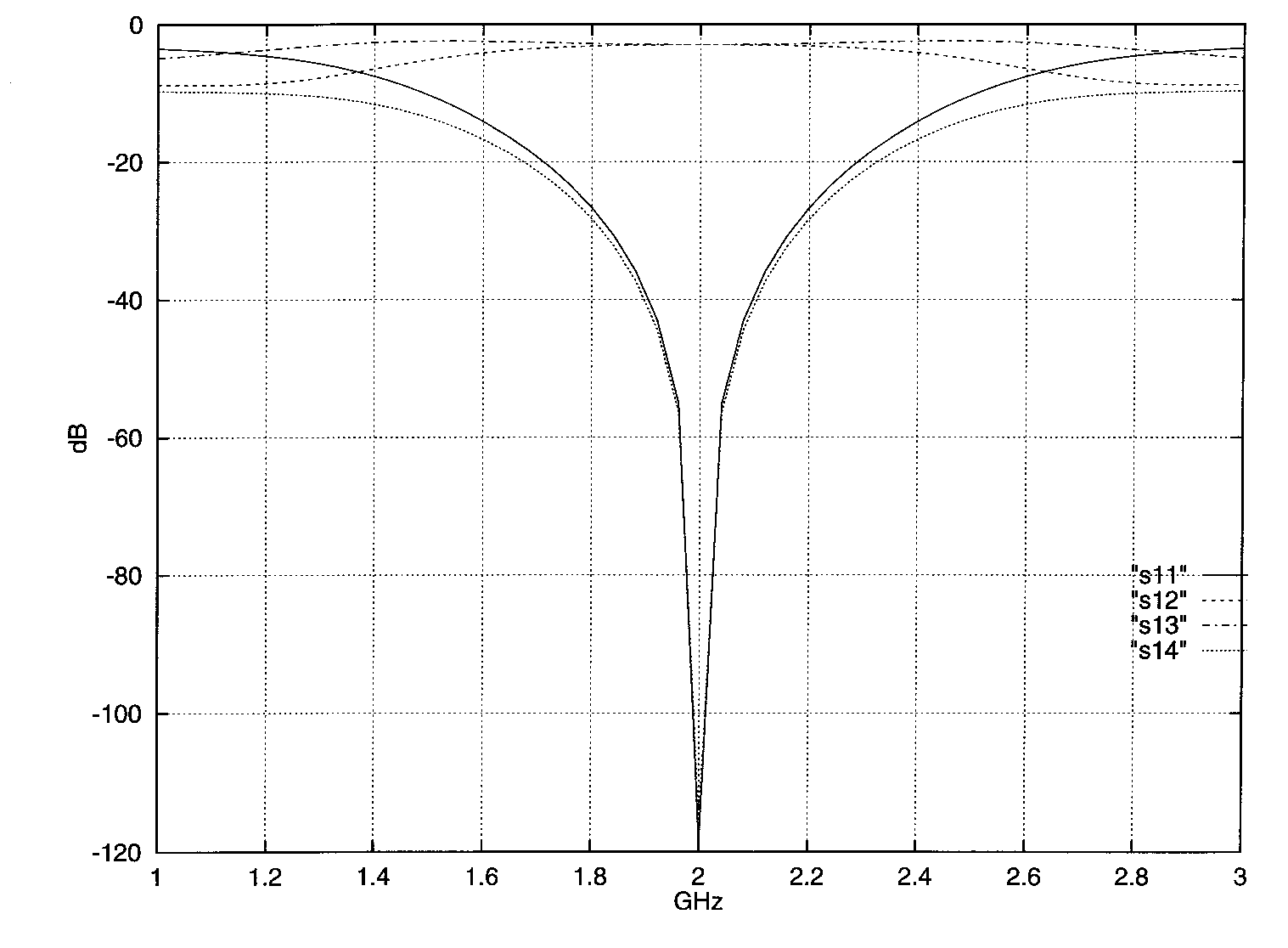}}
\caption{Computed response for the Butterworth synthesized
hybrid of Fig. 3. The hybrid is impedance transforming ($50 \Omega$
to $35\Omega $) and is required to have 0 dB power division ratio at
midband (2 GHz). The impedance values are $a_{1}= 129.20 \Omega,  a_{2}=
31.24\Omega, a_{3}= 79.37 \Omega, b_{1}= 33.56 \Omega, b_{2}= 27.55 \Omega$.}
\label{fig10}
\end{figure}

A  well-known synthesis method of two-port circuits is  the
Darlington method. In this method the response is specified
and the losslessness condition: $|\Gamma_{e}|^{2} + |T_{e}|^{2} =1$ is used to
extract  $|\Gamma_{e}|^2$.  The  next  step entails  extraction  of  a
complex  function $\Gamma_{e}$ from its modulus squared. In order  to
do  this extraction properly, the H\"{u}rwitz criterion must be
respected  [7].  Once  $\Gamma_{e}$ is obtained,  the  driving  point
impedance   of   the  circuit  $Z_{in}=(1+\Gamma_{e})/(1-\Gamma_{e})$   can   be
calculated.

The extraction of the shunt stubs and the DLUE from $Z_{in}$  is
done  sequentially. In the symmetric case  ($Z_{02}/Z_{01}=1$)  the
function  given by Levy et al. [6] and certified by  Riblet
[5]  was  used. However, we differ in the way we adapt  the
Darlington  synthesis to the extraction of  the  individual
elements.  For  instance, the formula used to  extract  the
first shunt stub $a_{1}$ is:

\begin{equation}
a_{1}=-\frac{1}{Z_{in}^2}{(\frac{dZ_{in}}{ds})}_{s=1} 
\end{equation}

where s is Richard's variable. The DLUE is extracted  by  a
sequential  extraction of two single length  unit  elements
(SLUE). A condition for the extraction of a SLUE is [7]:

\begin{equation}
Z_{in}(s=-1)=-Z_{in}(s=1)
\end{equation}

After  an  SLUE  is extracted, the driving point  impedance
becomes:

\begin{equation}
Z'_{in}(s)=Z_{in}(1) [\frac{sZ_{in}(1)-Z_{in}(s)}{sZ_{in}(s)-Z_{in}(1)}]
\end{equation}

For  the sequential extraction to work it is necessary that
the  transformed impedance satisfies (15). Further,  $Z_{in}(1)$
has to be equal to  $Z'_{in}(1)$ for the two extracted values  to
be  same.  This is a significant variation from the  method
used in [6]. Finally, the last shunt stub is extracted from
a  straight division of the denominator by the numerator of
the last $Z_{in}$.
In  the asymmetric case, we use the function $\Gamma_{e}/T_{e}$ given in
[4]  and proceed exactly as in the symmetric case. In  this
case $Z_{02}/Z_{01}= r $ where $r\ne 1$, the function  $\Gamma_{e}/T_{e}$ for an (n-1)-section hybrid is given by:

\begin{equation}
\frac{\Gamma_{e}}{T_{e}}=\frac{1}{\sqrt{r}}\frac{P_{n-1}(X/X_{c})}{P_{n-1}(1/X_{c})}[(r-1)
-jKtg(\theta)]
\end{equation}               

This function depends on two parameters $X_{c}$ and $ K$ that ought
to  be  determined from the coupling $(|b_{3}|^2)$ at the  center
frequency    and    the   directivity   ripple    bandwidth
specifications as explained below.
In the Butterworth case, $X_{c}=1$ and the polynomial function 
$P_{n-1}(X/X_{c})$ is given by $X^{(n-1)}$ with $X=(1+s^{2})/(1-s^2)$. 
Only one parameter ($K$) needs to be determined from the specification of a given value for
the  midband  coupling  (incidentally,  the  same  procedure
applies  for  specified  midband  power  division  ratio  or
isolation). The value of $K$ is numerically found as the  root
of the following equation:

\begin{equation}
{[20 \mbox{ log}_{10}(|T_{e}-T_{o}|/2)- |b_{3}|^2]}_{K}=0
\end{equation} 

In the Chebyshev case, the polynomial function  $ P_{n-1}(X/X_{c})$ is
given by [6]:

\begin{eqnarray}
&& P_{n-1}(X/X_{c})=(1+\sqrt{ 1-X_{c}^{2}})T_{n-1}(X/X_{c})/2-  \nonumber \\
&& \hspace{1.5cm}  (1-\sqrt{1-X_{c}^{2}}) T_{n-3}(X/X_{c})/2
\end{eqnarray}

where $ T_{n}(x)$ are the generalized Chebyshev functions defined
over the entire real axis.
The  Chebyshev case is more complex since one  has  to  find
numerically  the bandwidth parameter $X_{c}$ and the parameter  $K$
from the roots of the following coupled equations:

\begin{equation}
{[20 \mbox{ log}_{10}(|T_{e}-T_{o}|/2)- |b_{3}|^2]}_{K, X_{c}}=0
\end{equation}
and:

\begin{eqnarray}
&& {[20 \mbox{ log}_{10}\{|(T_{e}-T_{o})/(\Gamma_{e}-\Gamma_{o})|\}]}_{K, X_{c}} \nonumber \\
&& \mbox{} - {[20 \mbox{ log}_{10}\{|(T_{e}-T_{o})/(\Gamma_{e}-\Gamma_{o})|\}]}_{K, X_{2}} -20=0  \nonumber \\
\end{eqnarray}

The  reference  parameter $X_{2}$ corresponds  to  the  frequency
where  the directivity falls by 20dB from its value  at  $X_{c}$.
Once  the parameters $X_{c}$ and $K$ have been determined from  the
specifications,  one  proceeds to the determination  of  the
$a_{i}$'s and $b_{i}$'s.
We  are now in a position to make detailed comparisons with
the  measurements  and  optimization  as  well.  The  first
example  we  tackle  is the wideband two section  impedance
transformer  ($50\Omega$  to  $35  \Omega$)  with  specified  zero  power
division  at  midband.  Optimization and  measurements  are
compared  for  this hybrid in Figs 3 and 4 while  synthesis
results  are shown in Figs 9 and 10. We display all  the  S
parameters  for  both Butterworth and  Chebyshev  types  of
response.  The resulting impedance values in the  Chebyshev
case are: $a_{1}= 157.50 \Omega, a_{2}= 85.05 \Omega, a_{3}= 99.88 \Omega, b_{1}= 39.49
\Omega,  b_{2}=  33.49  \Omega$. When the specified type of  response  is
Chebyshev,  while a good isolation is obtained  at  midband
(-20 dB),  a  zero  power division  ratio  could  not  be
obtained.  This happens if a wide band, good isolation  and
impedance  transformation ratio of 0.7  are  simultaneously
required.  The  actual  power division  ratio  obtained  is
around  -5 dB. When any of these conditions is relaxed  the
required  solution exists and is shown in Fig. 10  for  the
Butterworth case and in Fig. 11 for the modified  Chebyshev
case as explained below.\\

\begin{figure}[htbp]
\centering{\includegraphics[width=2.5in,angle=0]{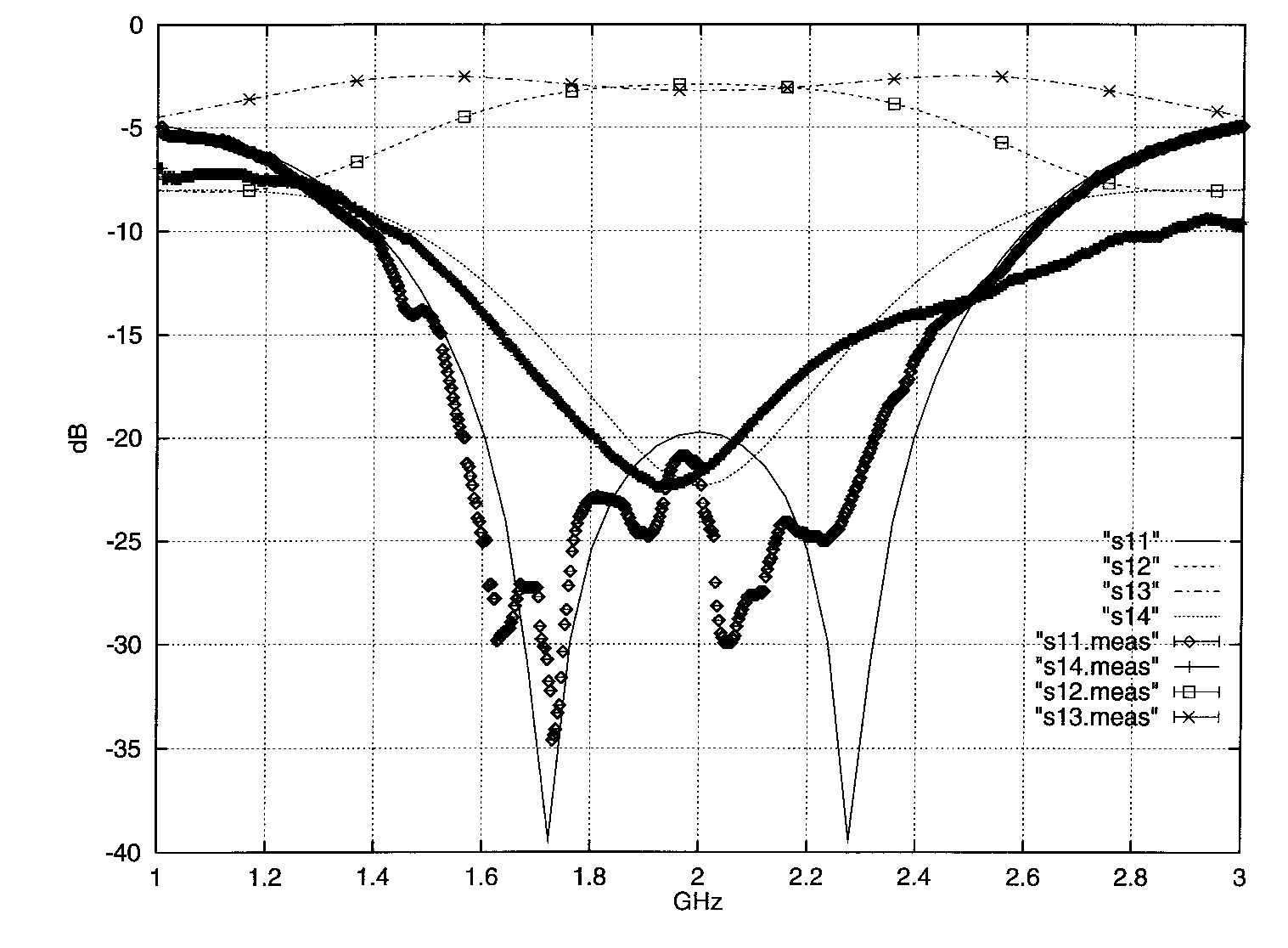}}
\caption{Computed and measured responses for the modified
Chebyshev  synthesized  hybrid.  The  constants  $K$  and   $X_{c}$
required  for the synthesis are calculated with specified  0
dB  power  division ratio at midband but with $r=1$. Once  the
synthesis is done, the S parameters are calculated from  the
cascaded  elements using the right value for  $r$ (0.7).  The
impedances are $a_{1}= 100 \Omega, a_{2}= 43 \Omega, 
a_{3}= 100 \Omega, b_{1}= b_{2}=  35.2\Omega$. 
The isolation obtained at midband is about -22.5 dB.}
\label{fig11}
\end{figure}

\begin{figure}[htbp]
\centering{\includegraphics[width=2.5in,angle=0]{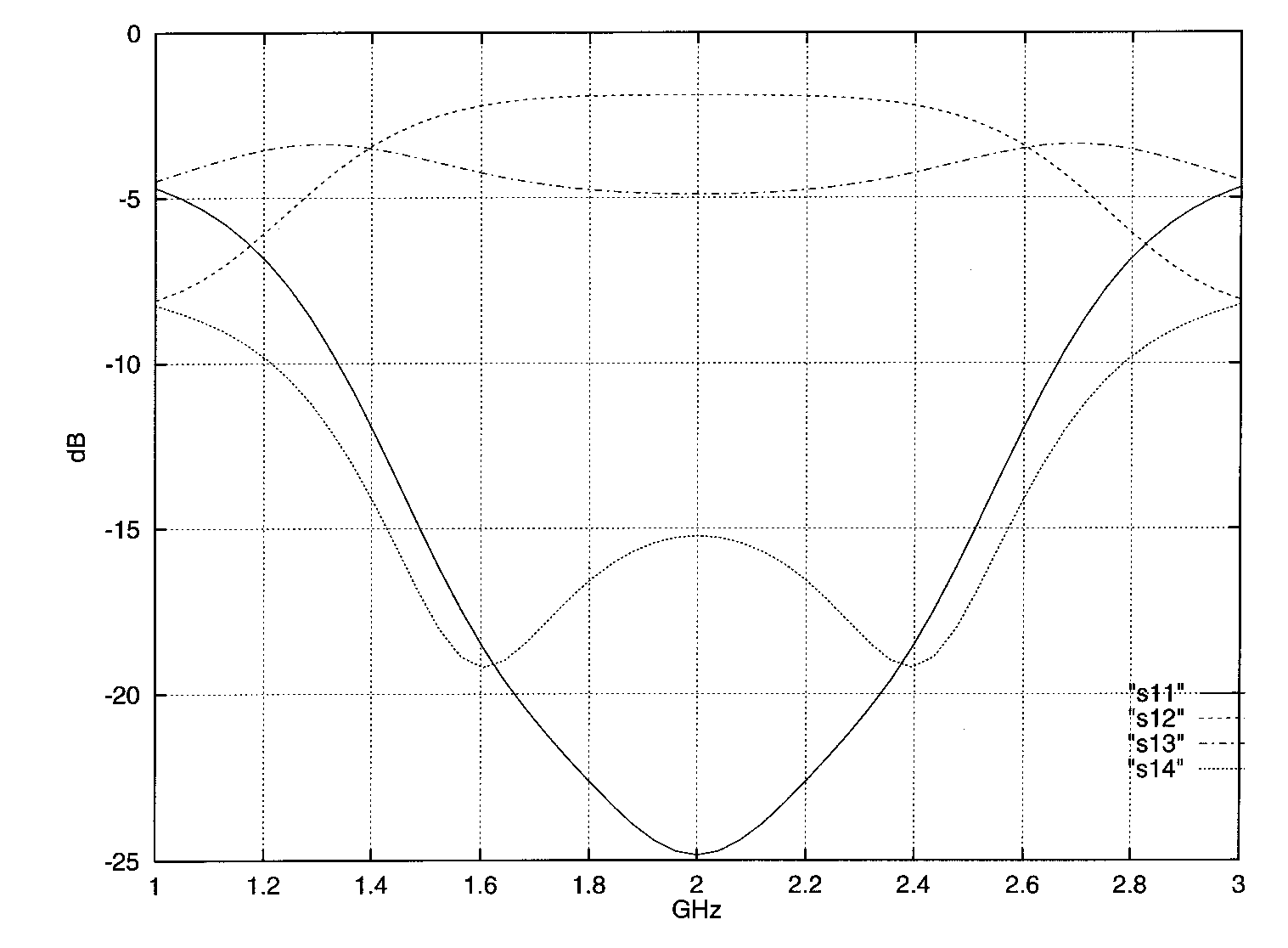}}
\caption{Computed response for the Chebyshev synthesized
3 dB   unequal  power  division  2 GHz  hybrid   ($r=1.2$)
corresponding to Fig. 7. The obtained isolation at  midband
is about -15 dB. The impedance values are $a_{1}= 161.31 \Omega, a_{2}=
55.34 \Omega, a_{3}= 125.94 \Omega, b_{1}= 39.97 \Omega, b_{2}= 36.28 \Omega$.}
\label{fig12}
\end{figure}

\begin{figure}[htbp]
\centering{\includegraphics[width=2.5in,angle=0]{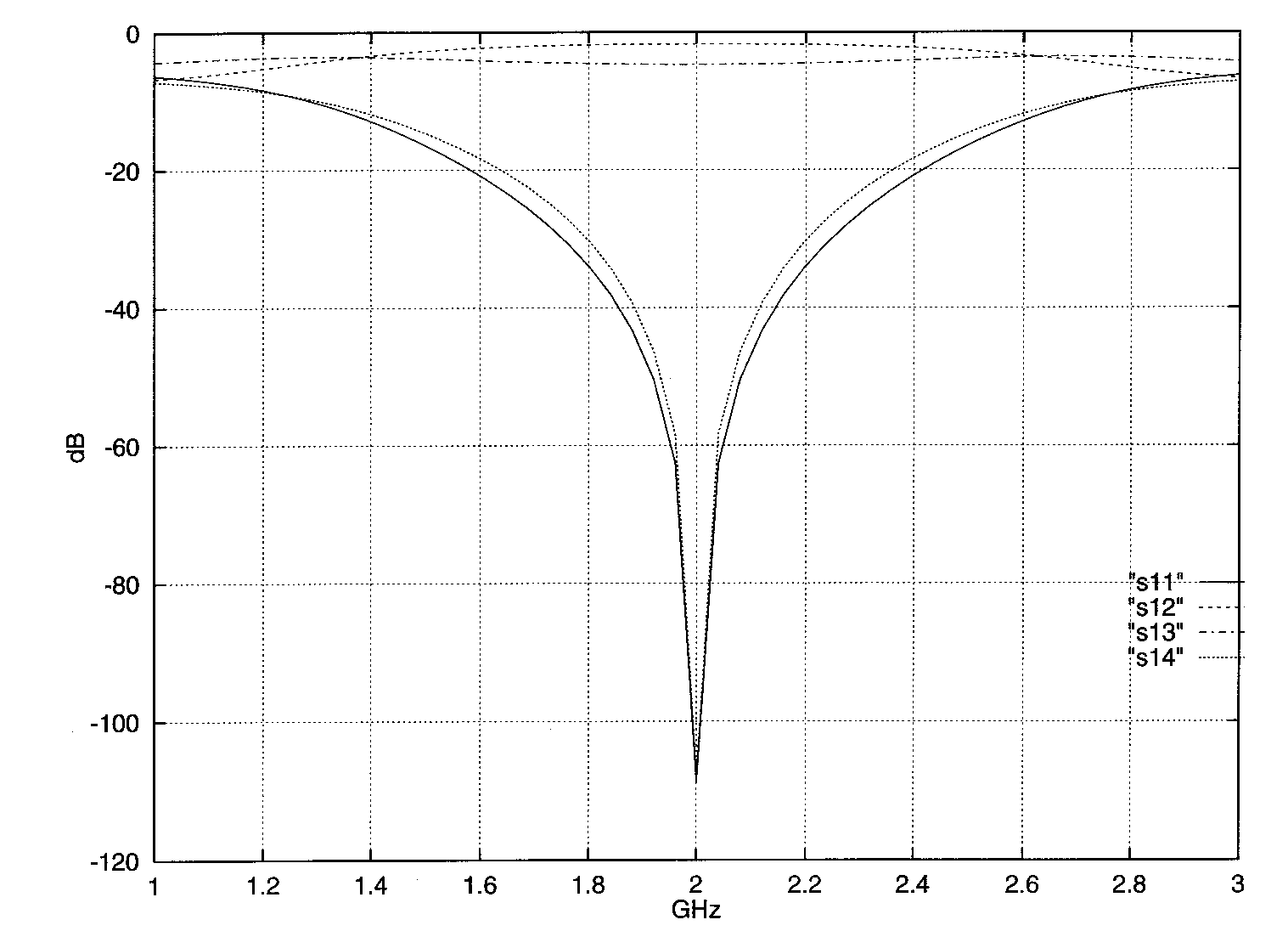}}
\caption{Computed response for the Butterworth synthesized 3 dB 
unequal power  division 2 GHz hybrid ($r=1.2$) corresponding to Fig. 7. 
The impedance values are $a_{1}=  152.44 \Omega, a_{2}= 66.05 \Omega,
 a_{3}= 195.31 \Omega, b_{1}= 43.55 \Omega,  b_{2}= 48.03 \Omega$.}
\label{fig13}
\end{figure}

\begin{figure}[htbp]
\centering{\includegraphics[width=2.5in,angle=0]{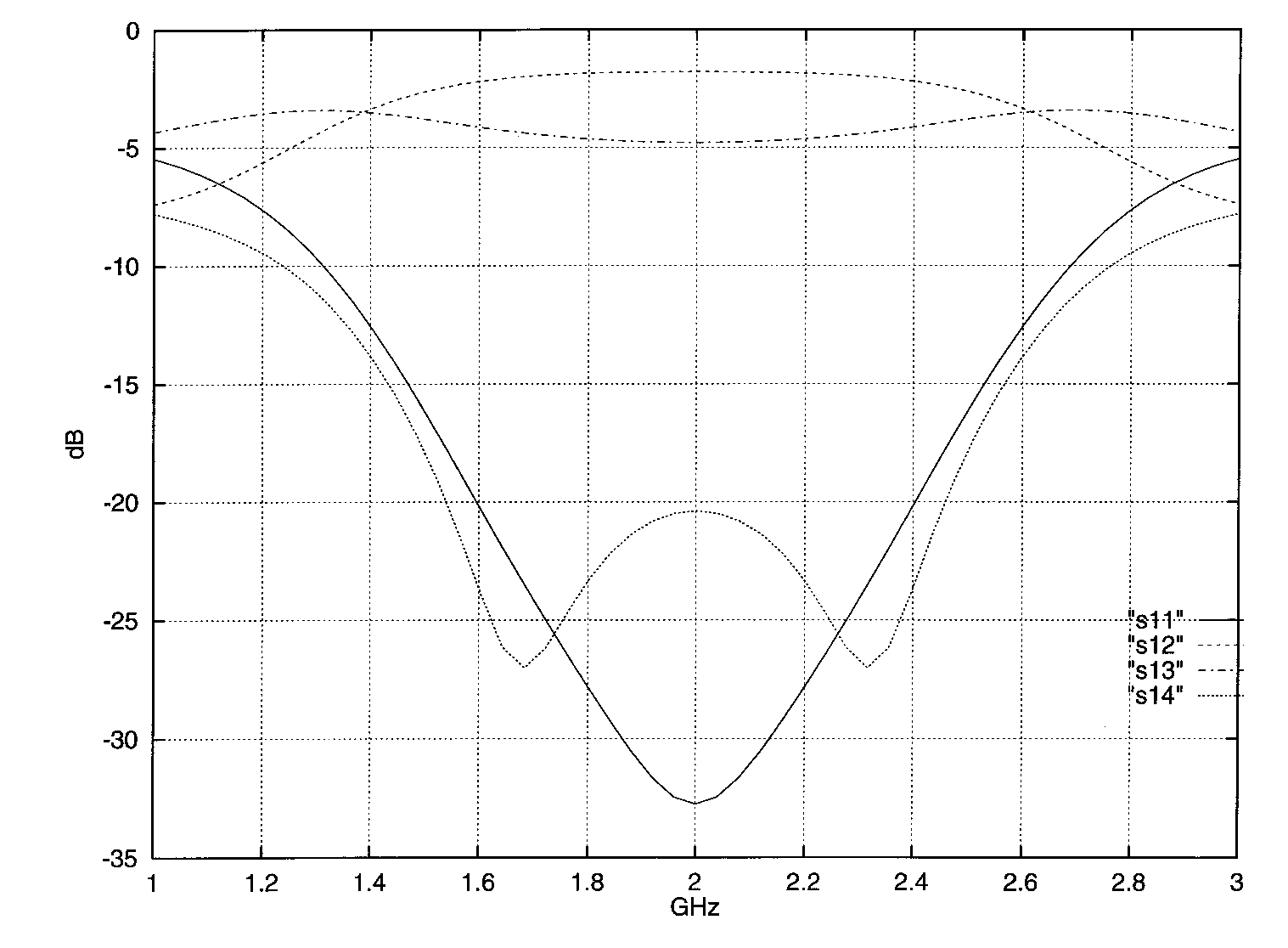}}
\caption{Computed response for the  modified  Chebyshev
synthesized  3  dB  unequal power  division  2  GHz  hybrid
($r=1.2$) corresponding to Fig. 7. The obtained isolation  at
midband  is less than -20 dB. The impedance values are $ a_{1}=
153.89 \Omega, a_{2}= 61.30 \Omega, a_{3}= 153.89 \Omega, b_{1}= b_{2}= 41.63 \Omega$.}
\label{fig14}
\end{figure}

The  modified Chebyshev approach consists of calculating the
parameters $K$ and $X_{c}$ by first assuming $r=1$. Once the $a_{i}$'s and
$b_{i}$'s are obtained, the S parameters with the actual value of
$r$  are  calculated. The frequency response is  displayed  in
Fig. 11. The required power division ratio (almost 0 dB)  as
well  as  good  isolation (around -22.5 dB)  were  obtained.
Experimental verification of the synthesized design was done
by  fabricating  a microstrip hybrid on a 0.031  inch  thick
Duroid substrate. The measured responses displayed in Fig.11
indicate a very close agreement with the synthesis.
The  second example of Chebyshev type is the wide  band  two
section impedance transformer ($50\Omega$ to 60 $\Omega$) with specified -
3dB power division at midband. This hybrid was optimized and
the  results were presented in Fig.7. In Fig. 12, all the  S
parameters   for  the  Chebyshev  synthesized   hybrid   are
presented. In this case, the resulting impedance values are:
$a_{1}=  161.31\Omega,  a_{2}=  55.37\Omega,  a_{3}= 125.94\Omega,  b_{1}=  39.96\Omega,  b_{2}=
36.28\Omega$.  The  power  division ratio was  obtained  from  the
synthesis as required (-3 dB) but a poor isolation (-15 dB)
at midband was found. In contrast, the Butterworth synthesis
is  shown  in  Fig.  13. Synthesis was also  done  with  the
modified  Chebyshev method mentioned above and  the  results
are shown in Fig. 14.
As  may  be seen, this modified method resulted in quite  a
reasonable response with an isolation better than 20dB.

\section{CONCLUSIONS}

Design  equations for a two section impedance  transforming
quad  hybrid  were  derived. Using these  equations  a  two
section  branch line hybrid can be designed  to  achieve  a
percentage  bandwidth of $30\%$ with impedance  transformation
by  a  factor of .7 to 1.3. Over this bandwidth  the  power
balance  between the output ports is measured  better  than
0.5  dB.  A two section branch line hybrid with 3dB unequal
power  division also has a $30\%$ bandwidth but the  impedance
transformation  ratio  range drops  to  [0.833  -  1.2].  A
slotline/lumped  implementation  of  such   a   hybrid   is
attractive  for MMIC circuits. In addition, a  new  general
synthesis method for a multisection hybrid with Butterworth
or  Chebyshev  response is described. Both symmetric  (with
equal   input  and  output  impedances)  and  non-symmetric
(impedance transforming) designs were demonstrated.
A  close  agreement between the synthesized, optimized  and
measured results were obtained.\\

{\bf Acknowledgement}

The  authors  wish  to  thank G. Wells  for  assistance  in
fabrication   and  testing  and  P.  Pramanick   for   many
stimulating  discussions. Financial support for  this  work
was  provided  by  the  Natural  Sciences  and  Engineering
Research  Council  of  Canada under a  university  industry
chair program.\\

\end{document}